\documentstyle[preprint,aps,epsf]{revtex}
\def\ao{\hat{a}}
\def\co{\hat{a}^\dagger}
\def\aop{\hat{a}^{\prime}}
\def\cop{\hat{a}^{\dagger\prime}}
\begin{document}
\draft
\title{Quantum State Tomography of Complex Multimode Fields using Array 
Detectors }

\author{S. Sivakumar$^{1,}$\footnote{Email: siva@igcar.ernet.in} and G.S. Agarwal $^{2}$}

\address{$^1$Indira Gandhi Centre for
Atomic Research, Kalpakkam 603 102 India\\
 $^2$Physical Research Laboratory, Navrangpura, Ahmedabad-
380009.\\ J.N. Center for Advanced Research, Bangalore,
India.  }
%\date{\today}
\maketitle
\begin{abstract}
    We demonstrate that it is possible
    to use the balanced homodyning with array detectors to measure the quantum
    state of correlated  two-mode signal field.  We  show the 
    applicability of the method to fields with complex mode functions, thus 
    generalizing the work of Beck(Phys. Rev. Lett. {\bf 84}, 5748 (2000)) in 
    several important ways. We further establish that,  under suitable 
    conditions, array detector measurements from one of the two ouputs is 
    sufficient to determine the quantum state of signal. We show
    the power of the method by reconstructing a truncated Perelomov state which
    exhibits complicated structure in the joint probability density for the
    quadratures.
\end{abstract}
\pacs{PACS Nos:42.50 Ar, 03.65 BZ.}
\section{Introduction}
The subject of quantum state tomography (QST) has been of great
interest in the recent years after the experimental reconstruction
of the complete wave function of the vacuum and squeezed states by
balanced homodyne (BHD) method\cite{smithey1,smithey2,breitenbach}. Much 
progress has been made by way of theoretical exploration of alternate schemes 
of QST like the self-homodyne tomography and their experimental
verification\cite{pkumar2}. In essence, the BHD method combines a
strong local oscillator (LO) with the signal in a 50:50
beam-splitter (BS). The two outputs of the BS are measured by two
photodetectors and electronically subtracted.  The resultant
quantity is directly proportional to the rotated quadrature
$[\ao\exp(-i\phi)+\co\exp(i\phi)]/\sqrt{2}$ of the signal field
and the angle $\phi$ is fixed by the LO.  By performing the
experiment many times for a given $\phi$ and repeating it for
various values of $\phi$ in the range of $0\le\phi\le \pi$, the
probability density $p(x,\phi)$ for the quadrature rotated through
$\phi$ can be measured.  From the measured probability density the
Wigner function for the signal can be constructed and hence the
elements of the density matrix\cite{risken,leonhardt}. 
The serious problems that may arise in the numerical reconstruction of density 
matrix elements and the methods to avoid such pitfalls are discussed in the 
review article by Welsch {\it et al} \cite{leonhardt} and in Ref. 
\cite{agarwal1}.  The BHD can be extended 
\cite{reymer1,reymer2,welsch,iaconis,agarwal2} to multimode signals   and to measure other distributions like the 
positive P-distribution.   The BHD is for optical fields and methods have been 
developed for reconstructing the vibrational state of molecules\cite{dunn} and 
spin systems\cite{agarwal4}.  For measruing the quantum states of the modes of 
a cavity, atoms with properly chosen energy levels can be used as probes
\cite{agarwal3}.
 
    A serious drawback of the BHD using single detectors is the
    requirement for {\it mode matching} the signal and LO mode functions.
     A quantitative measure of mode matching is the overlap
     between the signal and LO mode functions and the efficiency
     of the BHD scheme is directly proportional to it,
    
     \begin{equation}
     {\rm efficiency}\propto\int U^{*}_{\rm signal}(x)U_{\rm LO}(x)dx.
     \end{equation}
 Here $U_{\rm signal}$ and $U_{\rm LO}$ are the respective mode
     functions of the signal and LO.
     If there is perfect mode matching, {\it i.e.}, the overlap integral is
     unity, the BHD is very efficient.
In the case of signal and the LO modes being orthogonal to each
other, the BHD simply fails to determine the quantum state of the
signal.

An ingenious way of circumventing the problem of mode
 matching was suggested by Beck\cite{beck}.  The key idea is to replace
 the single detectors at the BS output ports by array
 detectors.  Beck considers array detectors  with pixels small enough
 such that the mode functions of the LO
 and signal over any pixel is constant.  The output from a
 given pixel labeled $(j,j')$ of one of the array detectors is given by
 \begin{eqnarray}
 N_{1}(j,j',\phi)={{\delta x\delta y
 c T}\over{2L}}\Big[{\vert\beta\vert^2\over{D_xD_y}}+
 \sum_{n,m}\co_{n}\ao_{m}U^*_{n}(j,j')U_{m}(j,j')\nonumber\\
 +{\beta\over{\sqrt{D_xD_y}}}\sum_m[\ao_m\exp(-i\phi)U_{m}(j,j')+H.c]\Big].
\end{eqnarray}
    Here the signal is taken to be multimode and the LO is assumed
to be a single mode coherent state $\vert\beta\exp^{i\phi}\rangle$
($\beta$ real). The summation is over the various modes of the signal field.
 The function $U_{m}(j,j')$ is the value of the $m$th mode function of the 
 signal over the pixel $(j,j')$.  The operators $\co_{m}$ and
 $\ao_{m}$ are the creation and annihilation operators for the $m$th mode of 
 the signal. The mode function for the LO is taken to be the constant
$1\over{\sqrt{D_x D_y}}$ over the entire detector surface.  The
dimension of each pixel is $\delta x\times\delta y$ and that of
the detector is $D_x\times D_y$ (=$N\delta x\times N\delta y$).
The constant $c$ is the speed of light vacuum, $T$ is the duration
of counting and $L$ is the longitudinal quantization length.
 Similarly, the output of the corresponding pixel at the other output
 port of BS is given by
\begin{eqnarray}
  N_{2}(j,j',\phi)&=&{{\delta x\delta y
  c T}\over{2L}}\Big[{\vert\beta\vert^2\over{D_xD_y}}+
  \sum_{n,m}\co_{n}\ao_{m}U^*_{n}(j,j')U_{m}(j,j')\nonumber\\
 &&-{\beta\over{\sqrt{D_xD_y}}} \sum_m[\ao_m\exp(-i\phi)U_{m}(j,j')+H.c]\Big].
\end{eqnarray}
On taking the difference of $N_{1}(j,j',\phi)$ and
$N_{2}(j,j',\phi)$, it is seen that
\begin{equation}\label{diff}
N_d(j,j',\phi)={{\delta x\delta y}\over{\sqrt{D_xD_y}}}\beta
\sum_m[\ao_m\exp(-i\phi)U_{m}(j,j')+H.c],
\end{equation}
where we have set $L=cT$.   If one of the mode functions, say
$U_l$, is real, then the above expression yields, on summing over
all the pixel indices and using orthonormal relationship among the
modes,
\begin{equation}\label{quadrature}
\sum_{j,j'}N_d(j,j',\phi)U_l(j,j')={1\over{\sqrt{D_x D_y}}}\beta
[\ao_l\exp(-i\phi)+\co_l\exp(i\phi)].
\end{equation}
 It is to be noted that the mode matching
factor does not enter in the expression relating the measured
photocurrent difference and the rotated quadrature.  This is of
great practical utility as mode matching is extremely difficult in
an actual experiment.  The above expression for quadrature works
for any mode function that is {\it real} and hence the quantum
state of all such modes can be determined.  However, their joint
state cannot be estimated by the method as proposed by Beck.  But correlated 
quantum states are of prime importance for conceptual understanding of
quantum theory as well as applications\cite{ou,bouwmeester}.
Hence, it is all the more essential that the {\it array detector
scheme be extended to correlated multimode fields whose mode
functions need not be real.}

    In the present  work  we extend the Beck's method to
    two-mode correlated states by interposing a linear device
    which introduces a variable relative phase between  the two
    modes.  The stringent condition to have real mode functions is relaxed
    by making measurements with the LO phase varying from zero to $3{\pi\over
    2}$ and combining the measurements to get the value of the quadratures 
    roatated through zero to $\pi$ as required for state reconstruction.
     Further, we establish that one of the outputs of BS as measured by an array
    detector  is sufficient to measure the quantum state.
    This paper is organized as follows.  In Section II the extension of Beck's
    work to two-mode correlated state is given. Section III of the paper
    contains the results of Monte Carlo simulations to demonstrate the power
    of the method to construct joint probability densities with complicated
    structures.  In Section IV of the paper we explicitly show that for
    mode functions satisfying certain conditions,
    a single array detector is sufficient to measure the quantum state.  We
    have provided an appendix to make transparent the arguments given
    Section II  for realxing the requirement for real mode functions.

\section{Joint quantum state of two-mode field}\label{joint}

  In the present section a
method of estimating the quantum state of a correlated two mode
signal is given.  The basic requirement is to construct the joint
probability density $P(x_1,\phi_1,x_2,\phi_2)$ for the two
quadratures rotated through $\phi_1$ and $\phi_2$ respectively. It
is, therefore, essential to rotate the two quadratures
independently.  If the signal modes are
spatially seperable, they can be mixed with two independent LOs (usually made
from a single source by using a BS) and
measurements are carried out using two BHD arrangements\cite{pkumar2}.  
In the other case,
the signal is passed through a medium which mixes the modes with a relative
phase. The mixed modes are taken to be spatially seperable\cite{reymer2} and one of the
seperated modes is used in the input port of BS.  This requires only one
LO and one BHD arrangment.  In both the cases, however, spatial seperation of
modes is
required at some stage and  mode matching is essential.
In the present section we discuss how to use the BHD with array detectors in the
latter method.
The mixing of modes is achieved by using a linear device which
introduces a relative phase between the two modes.  For instance,
the linear device could be a medium in which the two modes
interact {\it via} the Hamiltonian given by
\begin{equation}
H_{int}\propto \co_1\ao_2\exp(i\theta)+H.c
\end{equation}
The field operators $(\ao_1',\ao'_2)$ at the output of the linear
device can be written as
\begin{eqnarray}
\aop_1&=&{\cos\nu}\ao_1-i{\sin\nu}\exp(-i\theta)~\ao_2,\label{a1prime}\\
\aop_2&=&{\cos\nu}
\ao_2-i{\sin\nu}\exp(i\theta)~\ao_1,\label{a2prime}
\end{eqnarray}
where $\nu$ is a constant determined by the length of the linear
device, strength of interaction, etc.  The device mixes the two
modes of the signal.   For later use we define the rotated
quadratures for the transformed field:
\begin{equation}
\hat{X}'_{k}(\phi,\theta,\nu)={\aop_{k}\exp(-i\phi)+\cop_{k}\exp(i\phi)\over
\sqrt{2}},~~~~k=1,2
\end{equation}
  The positive frequency part
of the  electric field operator entering the signal port of  BS is
\begin{equation}
\hat{E}^{(+)^\prime}(x,y)=\Big[{2\pi\hbar\tilde{\omega}\over
L}\Big]^{1/2} \left[U_1(x,y)\aop_1+U_2(x,y)\aop_2\right].
\end{equation}
The electric field operator depends on $\phi$,$\theta$ and $\nu$
in addition to the spatial coordinates $(x,y)$.  However, in the
sequel we include the dependence explicitly in the expressions for
difference counts and other derived quantities.  With the field
operators given by Eq.\ref{a1prime} entering the signal port of BS
and the LO phase fixed at $\phi$,
 the difference count $N_d(j,j',\phi,\theta,\nu)$ is
\begin{eqnarray}
 N_d(j,j',\phi,\theta,\nu)&=&\beta{\delta x\delta
y\over{\sqrt{D_xD_y}}}\Big[\exp(-i\phi)[U_1(j,j')\aop_1+U_2(j,j')\aop_2]
\nonumber\\
& &+\exp(i\phi)[U_1^*(j,j')\cop_1+U_2^*(j,j')\cop_2]\Big].
\end{eqnarray}
and {\it the difference count with the  LO phase rotated further by
$\pi/2$ }is
\begin{eqnarray}
 N_d(j,j',\phi+{\pi\over 2},\theta,\nu)&=&i\beta{\delta x\delta
y\over{\sqrt{D_xD_y}}}\Big[\exp(i\phi)[U_1^*(j,j')\cop_1+U_2^*(j,j')\cop_2]
\nonumber\\
& &-\exp(-i\phi)[U_1(j,j')\aop_1+U_2(j,j')\aop_2] \Big].
\end{eqnarray}
The quantities $N_d(j,j',\phi,\theta,\nu)$ and
$N_d(j,j',\phi+{\pi\over 2},\theta,\nu)$ are measured in the
experiment. From these measured quantities we construct
    \begin{eqnarray}
    R(j,j',\phi,\theta,\nu)&=&{1\over{2\beta}} \sqrt{D_xD_y\over{2}}
    \{N_d(j,j',\phi,\theta,\nu)\nonumber\\
    &&{  }-i N_d(j,j',\phi+{\pi\over{2}},\theta,\nu)\},\\
    &=&{\delta x\delta y\over{\sqrt{2}}}\exp(i\phi)[\cop_1 U^*_1(j,j')+\cop_2
U^*_2(j,j')].
\end{eqnarray}
 Multiplying $R(j,j',\phi,\theta,\nu)$ by $U_k(j,j')$ $(k=1,2)$ and
summing over $(j,j')$ yields
\begin{eqnarray}
\sum_{j,j'}R(j,j',\phi,\theta,\nu)U_k(j,j')&=&{1\over{\sqrt{2}}}\exp(i\phi)
\cop_k.
\end{eqnarray}
The quadrature $\hat{X}{^\prime}_k(\phi)$ can now be written in
terms $R(j,j',\phi,\theta,\nu)$ and its conjugate as
\begin{eqnarray}
X{^\prime}_k(\phi)=\sum_{j,j'}[R(j,j{^\prime},\phi,\theta,\nu)U_k(j,j')+
C.c].
\end{eqnarray}
The equations determine the mixed quadratures in terms of measured
difference counts.
 Using Eqs. \ref{a1prime}-\ref{a2prime}, the quadratures for the mixed mode can 
 be written in terms of the signal quadratures as
\begin{eqnarray}
\hat{X}{^\prime}_{1}(\phi,\theta,\nu)&=&\cos\nu\hat{X}_1(\phi)+\sin\nu
\hat{X}_2(\phi+\theta+{\pi\over 2}),\label{x1prime}\\
\hat{X}{^\prime}_{2}(\phi,\theta,\nu)&=&\cos\nu\hat{X}_2(\phi)+\sin\nu
\hat{X}_1(\phi-\theta+{\pi\over 2}).\label{x2prime}
\end{eqnarray}
If measurements are carried out for two values of $\nu$, say
$\nu_1$ and $\nu_2$, then the signal quadratures can be evaluated
from Eqs.(\ref{x1prime})-(\ref{x2prime}) and the resulting
expressions are:
\begin{eqnarray}
\hat{X}_2(\phi)&=&{1\over{\sin(\nu_2
-\nu_1)}}[\sin\nu_1\hat{X}{^\prime}_2(\theta,\phi,\nu_2)-\sin\nu_2
\hat{X}{^\prime}_2(\theta,\phi,\nu_1)],\\
 \hat{X}_1(\phi-\theta+{\pi\over{2}})&=&{1\over{\sin(\nu_1
-\nu_2)}}[\cos\nu_1\hat{X}{^\prime}_2(\theta,\phi,\nu_2)-\cos\nu_2
\hat{X}{^\prime}_2(\theta,\phi,\nu_1)],\\
\hat{X}_1(\phi)&=&{1\over{\sin(\nu_2
-\nu_1)}}[\sin\nu_1\hat{X}{^\prime}_1(\theta,\phi,\nu_2)-\sin\nu_2
\hat{X}{^\prime}_1(\theta,\phi,\nu_1)],\\
 \hat{X}_2(\phi+\theta+{\pi\over{2}})&=&{1\over{\sin(\nu_1
-\nu_2)}}[\cos\nu_1\hat{X}{^\prime}_1(\theta,\phi,\nu_2)-\cos\nu_2
\hat{X}{^\prime}_1(\theta,\phi,\nu_1)].
\end{eqnarray}
The first two expressions yield the values of the quadratures
$\hat{X_2}$ and $\hat{X_1}$  rotated through angles $\phi$ and
$\phi-\theta+{\pi\over 2}$ respectively.  The last two yield the
value of two quadratures rotated through $\phi_1=\phi$ and
$\phi_2=\phi+\theta+{\pi\over{2}}$.
 By changing the LO phase
$\phi$ and the phase $\theta$ in the interaction Hamiltonian, the
quadratures can be measured to construct the joint probability
distribution $P(x_1,\phi_1,x_2,\phi_2)$.
Note that there is
{\it no need} to assume that the {\it mode functions are real}. Of
course, the difference count measurements are to be carried out
over a range of $0-3\pi/2$ for the LO phase; in the case of real
mode functions it is sufficient to measure the difference count
when the LO phase is varied from $0-\pi$.
 From the measured joint probability density, the Wigner function for the
two-mode state can be constructed which, in turn, can be used to
determine the density matrix. The measured joint probability can
also be used to construct the elements of the density matrix in
two-mode number state basis\cite{dariano2}.

The scheme suggested here is similar to what is described as the
method of generalized rotation in phase space\cite{reymer2}.
However, in our scheme there is no need to spatially separate the
signal modes at the output of the linear device.  This is possible
as mode matching is not essential in the present scheme.   In
fact, not separating them is useful to reduce the number of
experimental runs.  In one run of the experiment, the pairs
$\{\hat{X}_1(\phi),\hat{X}_2(\phi+\theta+{\pi\over
 2})\}$ and $\{\hat{X}_1(\phi-\theta+{\pi\over
2}),\hat{X}_2(\phi)\}$ are simultaneously estimated.

\section{Monte Carlo Simulations}

    In this section we study the experimental {\it feasibility} of the
    above method by Monte Carlo (MC) simulations with the signal in the
    two-mode Perelomov state\cite{caves}.  This state is generated from the 
    two-mode vacuum $\vert 0,0\rangle$ as follows:
    \begin{equation}
    \vert
    \zeta\rangle=\exp(\zeta\co_1\co_2-\zeta^*\ao_1\ao_2)\vert
    0,0\rangle.
\end{equation}
The number state expansion for the above state is given by
\begin{equation}
\vert \zeta\rangle ={1\over{\cosh
r}}\sum_{n=0}^{\infty}[\exp(-i\gamma)\tanh r]^n\vert n,n\rangle,
\end{equation}
and the joint probability density is
\begin{equation}\label{probdensity}
p(x_1,\phi_1,x_2,\phi_2)={2\over {\pi {\rm A B}}}
\exp\Big[-{(x_1+x_2)^2\over{\rm A}}-{(x_1-x_2)^2\over{\rm B}}\Big].
\end{equation}
The constants $r$, $\rm A$ and $\rm B$ are related to the squeeze
parameter $\zeta$:
\begin{eqnarray}
\zeta&=&r\exp(-i\gamma),\nonumber\\ A&=&{\vert 1+\tanh
r\exp[-i(\phi_1+\phi_2+\gamma)]\vert^2\over{1-\tanh^2r}}\nonumber\\
B&=&{\vert 1-\tanh r
\exp[-i(\phi_1+\phi_2+\gamma)]\vert^2\over{1-\tanh^2r}}\nonumber;
\end{eqnarray}

Random numbers distributed according to the distribution given in Eq.
\ref{probdensity}  were
generated from Gaussian distributed random numbers by von
Neumann's rejection method\cite{numreci}. These numbers were then
used to generate the output of an actual experiment.  In Fig. 1
the result of the simulation to reconstruct the joint probability
density is given for $\phi_1=\pi/4$, $\phi_2=\pi/2$,
$\gamma=\pi/4$ and $r=1.0$ along with its contour plot . For
comparison, the theoretical distribution is also given. We find
that to reconstruct the probability density with reasonable
accuracy, as many as $1.6\times 10^5$ experimental runs would be
required for each set of values of $\phi_1$ and $\phi_2$. In Fig.
2 we present the simulated results for identical values for the
parameters except $\phi_2$ which is set equal to $-\pi/4$. In Fig. 1
the distribution is narrower compared to what is shown in Fig, 2.
As expected, from the
figures it is clear that the reconstruction is better, for a given number of
experiments, if the probability density is narrower.

    In order to see whether an experiment can capture the
    {\it complicated spatial structures} of probability density, the MC
    simulations were done for the signal in truncated Perelomov
    state.  These are defined as
    \begin{equation}
    \vert c_1,c_2\rangle=c_1\vert 0,0\rangle + c_2\exp(-i\delta)
    \vert 1,1\rangle.
    \end{equation}
The superposition coefficients $c_1$ and $c_2$ are taken to be real and 
they satisfy the
normalization condition $c_1^2+c_2^2=1$. The joint probability
density for the quadratures is
\begin{eqnarray}
p(x_1,\phi_1,x_2,\phi_2)={\exp{(-x_1{^2}-x_2{^2})}\over\pi}[
c_1^2+4c_2^2 x_1{^2} x_2{^2}+4x_1 x_2 c_1
c_2\cos(\phi_1+\phi_2+\delta)].
\end{eqnarray}
The result of the MC simulation for the truncated Perelomov state
is depicted in Fig. 3.  We have used $c_1=c_2={1/\sqrt 2}$,
$\delta=\pi/8$, $\phi_1=\pi/4$ and $\phi_2=\pi/4$.  It requires
$2\times 10^5$ points to generate the pattern shown.

\section{ QST using a single array detector}

    In this section we describe how to measure the joint
    probability density of a two-mode state using an array
    detector in one of the output ports of BS. No measurements are
    made at the other output.  In the discussions
    to follow the pixels are labeled by a single index instead of
two indices.  We first introduce some
    notations to simplify the expressions:
    \begin{eqnarray}
    U_1(k) &:& \textrm{value of the mode function $U_1$ on the
    pixel labeled $k$},\nonumber\\
    U_2(k) &:&\textrm{value of the mode function $U_2$ on the
    pixel labeled $k$},\nonumber\\
    n_d(k) &:& \textrm{ difference count between the pixels labeled $1$ and
    $k$\quad $(k=2,3...9)$}\nonumber\\
    V^T &:& \textrm{transpose of vector $V$ whose elements are}\nonumber\\
  && \{\cop_1\aop_1,\quad\cop_2\aop_2,\quad\cop_1\aop_2,
  \quad\aop_1\cop_2,\quad\exp(-i\phi)\aop_1,\quad\exp(i\phi)\cop_1,\nonumber\\
    && \quad\exp(-i\phi)\aop_2,\quad\exp(i\phi)\cop_2\}\nonumber\\
    M &:& \textrm{an $8\times 8$ matrix whose $(k-1)$th $(k=2,..9)$ row is 
    given by}\nonumber\\
    &&\{ U^2_1(1)-U^2_1(k),\quad U^2_2(1)-U^2_2(k),\quad U^*_1(1)U_2(1)-
    U^*_1(k)U_2(k),\nonumber\\
    &&U_1(1)U^*_2(1)-U_1(k)U^*_2(k),\quad S[U_1(1)-U_1(k)],\nonumber\\&&
    S[U^*_1(1)-U^*_1(k)],\quad S[U_2(1)-U_2(k)],\quad
    S[U^*_2(1)-U^*_1(k)]\},\nonumber\\& &{ }\textrm{where}~~
    S={\beta\over\sqrt{D_xD_y}}\nonumber
    \end{eqnarray}

The expression for $n_d(k)~(k=2,3\cdot\cdot\cdot 9)$ is
\begin{eqnarray}
n_d(k) & = & \textrm {Counts in pixel 1 - Counts in pixel
$k$}\nonumber\\ & = &{\delta x\delta y\over
2}\Big[\sum_{n,m=1}^2\cop_n\aop_m
[U^*_n(1)U_m(1)-U^*_n(k)U_m(k)]\nonumber\\ &&{  }+
{\beta\over{\sqrt{D_xD_y}}}\sum_m[\exp(-i\phi)\aop_m[U_m(1)-U_m(k)]+
H.c]\Big]\\ & = &{\delta x\delta y\over 2}\sum_{j=1}^8M_{kj}V_j.
\end{eqnarray}
The elements of the matrix are various combinations of the mode
functions $U_1$ and $U_2$ over a selected set of nine pixels of
the array detector.  If the elements of $M$ are such that the
matrix is invertible, the elements of the vector $V$ can be
determined from the equation
\begin{equation}
V_j=\sum_{k=2}^9 (M^{-1})_{j,k-1}n_d(k)\quad j=1,2\cdot\cdot\cdot
8.
\end{equation}
 Note that the vector $V$ has as its elements the creation and
annihilation operators of the two modes and their quadratic forms. The above
equation then implies that the elements of $V$ are determined from
the measured quantities $n_d(k)$. It is evident that the mixed
quadratures can be determined as
\begin{eqnarray}
\hat{X}{^\prime}_1(\phi,\theta,\nu)&=&{V_5+V_6\over\sqrt 2},\\
\hat{X}{^\prime}_2(\phi,\theta,\nu)&=&{V_7+V_8\over\sqrt 2}.
\end{eqnarray}
Once the mixed quadratures are estimated, the signal mode
qudratures are obtained using Eqs.\ref{x1prime}-\ref{x2prime}.   
It is clear that if we can choose nine pixels so that the matrix
$M$ is invertible, then the joint probability density of two-mode
states can be determined.  Despite the fact that  only one of the
output ports of BS is used, the LO fluctuations are eliminated as
in the case of BHD.  Another interesting feature is that there is
no need to assume that the mode functions are real and the LO
phase need not be changed beyond the usual range $0-\pi$. The
method can be extended to more than two modes.  But the
requirements on the mode functions will become more stringent.

\section{Summary}

    The use of array detectors in BHD eliminates the
need for difficult-to-achieve mode-matching.    For two-mode
signals, a linear device to introduce a relative phase between two
modes can be used in the signal port of the BS.  By varying the LO
phase from zero to $3\pi\over 2$ and the relative phase between
the modes, the joint probability density for the quadratures of
the two modes can be experimentally determined.  The QST
measurements can carried out with one array detector instead of
detectors at both the output ports of BS.   The method retains the
advantage of BHD in the sense that the LO fluctuations do not
affect the measurement.  However, the success of the method
depends of the shape of the mode functions.

\appendix
\section{QST of a single, complex mode signal}
In this appendix we show how to measure the quantum state of a single,
complex  mode signal.  This appendix is provided to make the arguments of
Section \ref{joint} more transparent.   In Beck's method it is required to have
the mode function real for determining its quantum state.  If the mode function
 is not real, then it cannot be factored out in the RHS of Eq. (\ref{diff}) and
 hence the measured difference count cannot be related to the quadrature.  Note
 that this problem does not arise in the case of the conventional BHD using
 single detectors.  We set $N_d(j,j',\phi)$ as
 $N_d(\phi)$ and $U(j,j')$ as $U$  to avoid lengthy expressions.
 Hence, Eq. (\ref{diff}) is rewritten as
 \begin{equation}
N_d(\phi)={{\delta x\delta y}\over{\sqrt{D_xD_y}}}\beta
[U\exp(-i\phi)\ao+U{^*}\exp(i\phi)\co].
\end{equation}
With the LO phase rotated through
 $\phi+{\pi\over 2}$, the difference count is
 \begin{equation}\label{eq1ofapp}
N_d(\phi+{\pi\over 2})=i{{\delta x\delta y}\over{\sqrt{D_xD_y}}}\beta
[\co\exp(i\phi)U{^*}-U\exp(-i\phi)\ao].
\end{equation}
Comibining the expressions for $N_d(\phi)$ and 
$N_d(\phi+{\pi\over 2})$ we get
\begin{equation}
N_d(\phi)-iN_d(\phi+{\pi\over 2})= U{^*}\exp(i\phi)\co
\end{equation}
and
\begin{equation}
N_d(\phi)+iN_d(\phi+{\pi\over 2})= U\exp(-i\phi)\ao.
\end{equation}
Having measured the operators $\exp(-i\phi)\ao$ and its adjoint, the quadrature
can be estimated at once as it is a linear combination of the two operators.
To achieve this
 we multiply the last two equations by $U$ and $U{^*}$ respectively  and
 use the normalization relation $\delta x\delta y\sum_{j.j'}U^*U=1$ to yield
 \begin{equation}
 \hat{X}(\phi)={\sqrt{D_x D_y}\over{2\sqrt{2} \beta}}
 \sum_{j,j'}[N_d(\phi)(U+U^*)+iN_d(\phi+{\pi\over 2})(U^*-U)].
 \end{equation}
As asserted in the beginning of the appendix, there is no need to
assume the mode functions to be real in the above expression. When
$U$ is real the second term on the RHS vanishes and we recover, as
expected, Eq.\ref{quadrature}.

\newpage
\begin{figure}[h]
\hspace*{1.5 cm} \epsfxsize=400pt \epsfbox{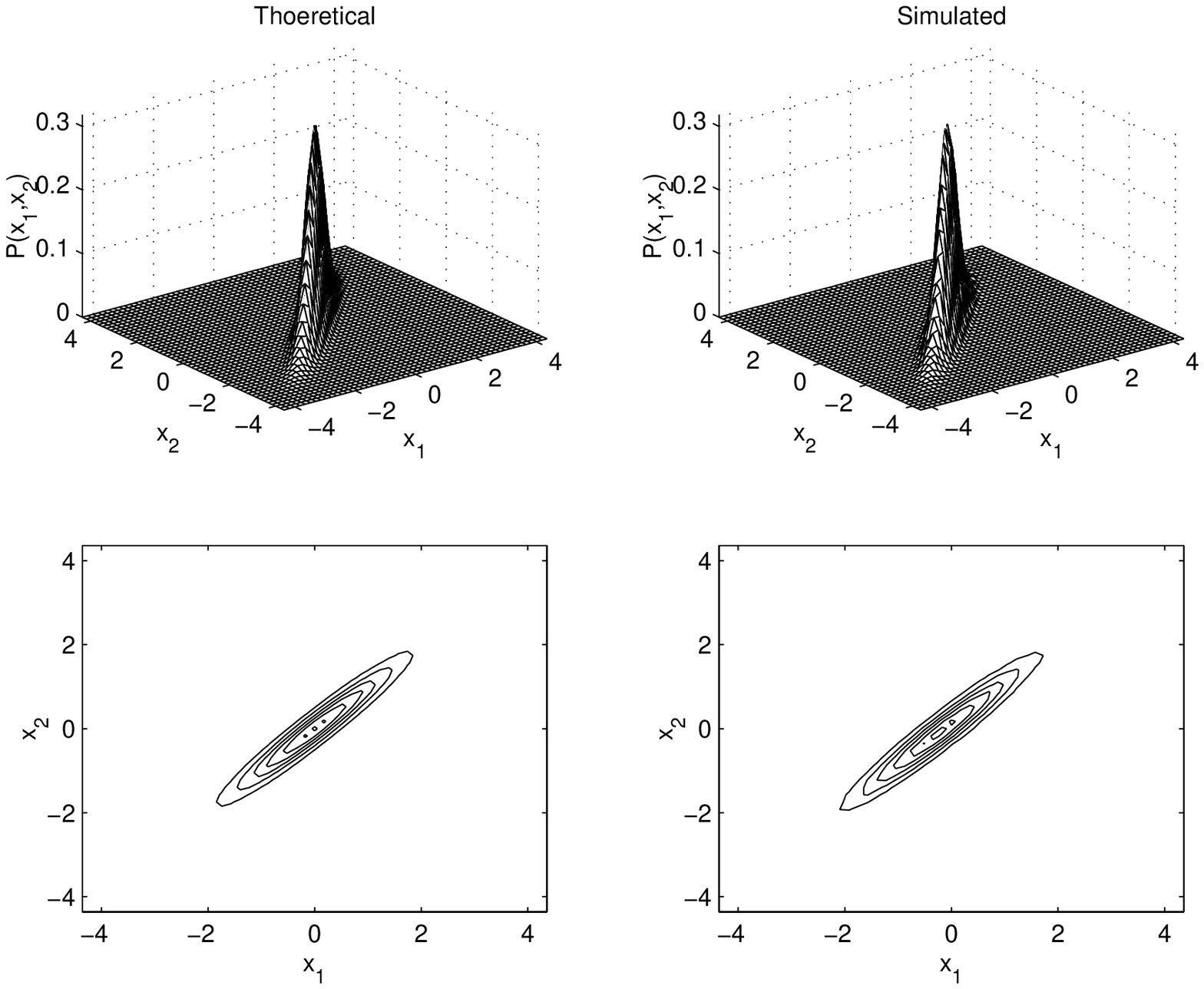} \vspace*{2
cm} \caption{ Joint probability density for two-mode Perelomov
state. The parameter values are $\phi_1=\pi/4$, $\phi_2=\pi/2$, $\gamma=\pi/4$
and
$ r=1.0$. } \label{fig1}
\end{figure}
\newpage
\begin{figure}[h]
\hspace*{1.5 cm} \epsfxsize=400pt \epsfbox{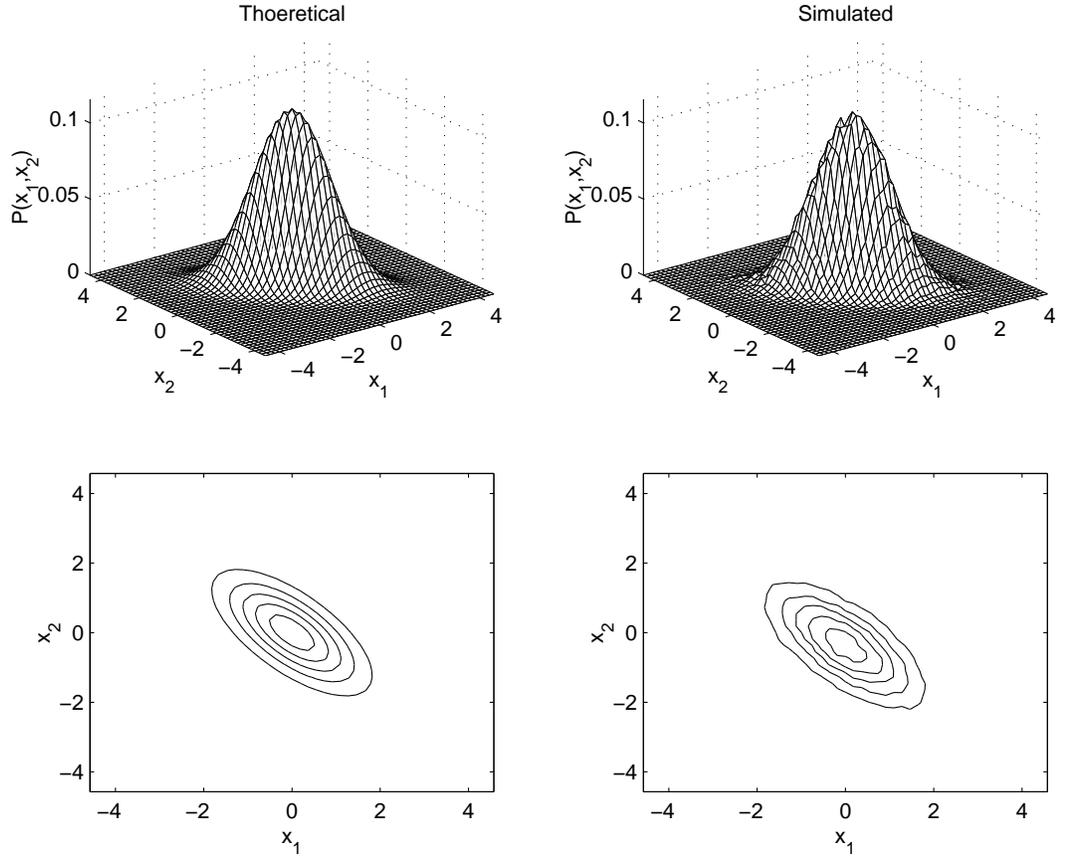} \vspace*{2
cm} \caption{Joint probability density for two-mode Perelomov
state with with the parameters taking the same values as in Fig. 1
except $\phi_2$ which is set equal to $-\pi/4$.} \label{fig2}
\end{figure}

\newpage
\begin{figure}[h]
\hspace*{1.5 cm} \epsfxsize=400pt \epsfbox{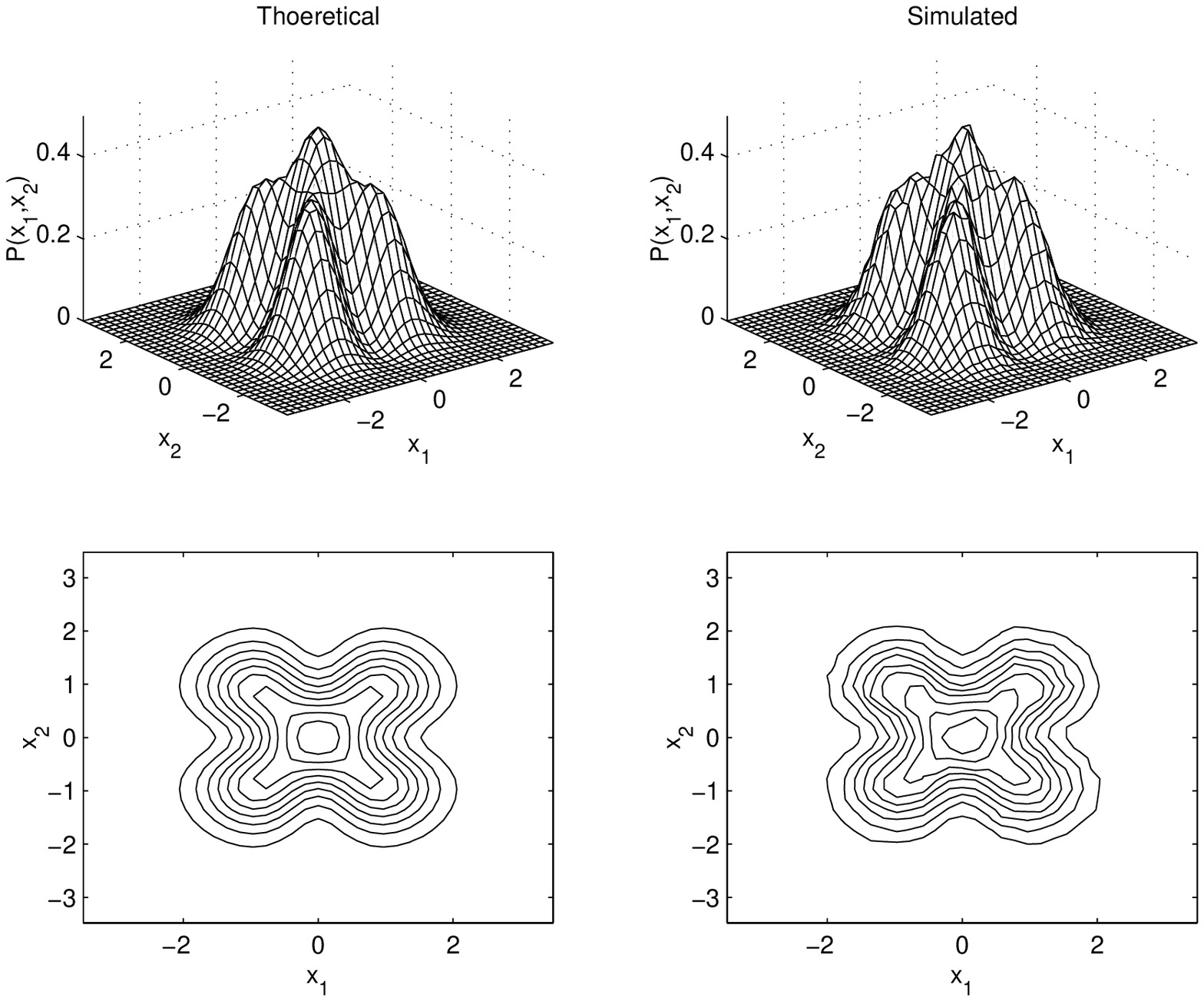} \vspace*{2
cm} \caption{ Theoretical and simulated joint probability
densities for the truncated Perelomov state.  The
parameter values are $c_1=c_2=1/\sqrt{2}$  and $\delta=\pi/8$}
\label{fig3}
\end{figure}

\begin{thebibliography}{999}

\bibitem{smithey1}{D.T. Smithey, M.Beck, M.G. Raymer, and A.
Faridani, Phys. Rev. Lett. {\bf 70}, 1244 (1993).}
\bibitem{smithey2}{D.T.Smithey, M. Beck, J. Cooper, and M.G. Raymer, 
Phys. Rev. A {\bf
48}, 3159 (1993).}
 \bibitem{breitenbach}{ G. Breitenbach, T. Muller, S.F. Pereira, J.-Ph.
Poziat, S. Schiller, and J. Mlynek, J. Opt. Soc. Am. B, {\bf 12},
2304 (1995).}
\bibitem{pkumar2}{M. Vasilyev, S.-K. Choi, P. Kumar and G.M. D'Ariano,
Phys. Rev. Lett. {\bf 84}, 2354 (2000).}
\bibitem{risken}{K. Vogel, and H. Risken, Phys. Rev. A {\bf 40}, 2847 (1989).}
\bibitem{leonhardt}{For review see: U. Leonhardt, {\it Measuring the Quantum 
State of Light}, Oxford University Press, Oxford (1997);  D.-G. Welsch, W.
Vogel, and  T. Opantry,{\it Progress in Optics}, Ed. E. Wolf, Vol.
{\bf XXXIX} (1999).}
\bibitem{agarwal1}{G. S. Agarwal and J. Banerjee, LANL e-print quant-ph 
0007049 (submitted to Phys. Rev. A.)}
\bibitem{reymer1}{M.G. Raymer, D.F. McAlister, and U. Leonhardt,
Phys. Rev. A. {\bf 54}, 2397 (1996).}
\bibitem{reymer2}{M.G. Raymer, and A.C. Funk, Phys. Rev. A. {\bf
61}, 015801 (1999).}
\bibitem{welsch}{T. Opantry, D.-G. Welsch, and W. Vogel, Phys.
Rev. A {\bf 55}, 1416 (1997).}
\bibitem{iaconis}{C. Iaconis, E. Mukamel, and I. A. Walmsley J. Opt. B {\bf 2},
510 (2000).}
\bibitem{agarwal2}{G. S. Agarwal and S. Chaturvedi, Phy. Rev. A {\bf 49},
R665(1994).}
\bibitem{dunn}{T. J. Dunn, J. N. Sweetser, I. A. Walmsley, and C. Radzewicz,
Phys. Rev. Lett. {\bf 70}, 3388 (1993); T. J. Dunn, I. A. Walmsley, and S.
Mukamel, {\it ibid.} {\bf 74}, 884 (1995).} 
\bibitem{agarwal4}{G. S. Agarwal, Phy. Rev. A {\bf 57}, 671(1998).}
\bibitem{agarwal3}{M. S. Kim and G. S. Agarwal, Phy. Rev. A {\bf 59}, 
3044 (1999).}
\bibitem{beck}{M. Beck, Phys. Rev. Lett. {\bf 84}, 5748 (2000).}
\bibitem{ou}{Z. -Y. Ou and L. Mandel, Phys. Rev. Lett {\bf 61}, 50
(1988); Y. H. Shih and C. O. Alley, Phys. Rev. Lett. {\bf 61}, 2921
(1988).}
\bibitem{bouwmeester}{D. Bouwmeester, J. -W. Pan, K. Mattle, M.
Eibl, H. Weinfurter, and A. Zeilinger, Nature (London) {\bf 390},
575 (1997); D. Boschi, S. Branca, F. De Martini, L. Hardy, and S.
Popescu, Phys. Rev. Lett. {\bf 80}, 1121 (1998); A. Furusawa, J.
L. Sorensen, S. L. Braunstein, C. A. Fuchs, H. J. Kimble, and E.
S. Polzik, Science {\bf 282}, 706 (1998).}
\bibitem{dariano2}{G. M. D'Ariano, U. Leonhardt, and H. Paul,
Phys. Rev. A {\bf 52}, R1801 (1995).}
\bibitem{caves}{B. Schumaker and C. M. Caves, Phys. Rev. A {\bf 31}, 3093 
(1985).}
\bibitem{numreci}{W. H. Press, S. A. Teukolsky, W. T. Vettering, and
B. P. Flannery, {\it Numerical Recipes in FORTRAN}, Cambridge
University Press (1995).}
\end{thebibliography}
\end{document}